\documentclass[twoside]{article}\usepackage[margin=1in]{geometry}
 \newcommand\aut {L.A. Levin {\rm and} R. Venkatesan}
 \newcommand\ttl {Average Case NP-complete Graph Coloring Problem}
 \usepackage{amsthm}
 \usepackage {graphicx,amsmath,amssymb,cmap}
 \usepackage[unicode]{hyperref} 
\begin{document}

 \newtheorem{lem} {Lemma} \newtheorem {thm} {Theorem}
 \newtheorem{prop} {Proposition} \newtheorem{cor} {Corollary}
 \newtheorem {defn} {Definition} \newtheorem {rem} {Remark}

 \newcommand\BD {\begin{defn}} \newcommand\ED {\end{defn}}
 \newcommand\BL {\begin{lem}} \newcommand\EL {\end{lem}}
 \newcommand\BT {\begin{thm}} \newcommand\ET {\end{thm}}
 \newcommand\BP {\begin{prop}} \newcommand\EP {\end{prop}}
 \newcommand\BR {\begin{rem}\em } \newcommand\ER {\end{rem}}
 \newcommand\BC {\begin{cor}} \newcommand\EC {\end{cor}}
 \newcommand\BPR {\begin{proof}} \newcommand\EPR {\end{proof}}

 \newcommand\emm [1]{{\ensuremath{#1}}} \newcommand\drop [1]{}
 \newcommand\ov {\overline} \newcommand\pp[1] {\paragraph{\bf#1}}
 \newcommand\hreff [1]{{\small\url{http://#1}}}
 \newcommand\trm [1]{{\bf\em #1}}
 \newcommand\ceil [1]{{\lceil #1\rceil}}
 \newcommand\floor [1]{{\lfloor #1\rfloor}}
 \newcommand\edf {{\emm{\stackrel{\text{\tiny def}}=}}}

 \renewcommand\a {\alpha} \renewcommand\l {\lambda}
 \newcommand\w {\omega} \newcommand\W {\Omega} \newcommand\s {\sigma}
 \newcommand\e {\varepsilon} \newcommand\dgr {{\Delta}}
 \newcommand\Q {{\mathbb{Q}}} \newcommand\R {{\mathbb{R}}}
 \newcommand\Z {{\mathbb{Z}}} \newcommand\N {{\mathbb{N}}}
 \newcommand\St {{\mathbb{S}}} \newcommand\E {{\mathbb{E}}}
 \newcommand\Pb {{\mathbb{P}}}

 \newcommand\ic {{\mbox{\bf I}}} \newcommand\sX {{\scriptscriptstyle X}}
 \newcommand\key {{\emm\kappa}} \newcommand\dwn {{\mbox{\bf c}}}

 \newcommand\vl {{\mbox{\bf v}}} \newcommand\hl {{\mbox{\bf h}}}
 \newcommand\il {{\mbox{\bf i}}} \newcommand\jl {{\mbox{\bf j}}}
 \newcommand\kl {{\mbox{\bf k}}} \newcommand\lln {{\mbox{\bf l}}}
 \newcommand\lr {{\mbox{\bf s}}} \newcommand\bit {{\mbox{\bf b}}}
 \newcommand\fl {{\mbox{\bf f}}} \newcommand\wl {{\mbox{\bf w}}}
 \newcommand\trit {{\emm\tau}}

 \newcommand\del {{\emm{\scriptstyle\Delta}}}
 \newcommand\CC [1]{{\stackrel{ #1}{\longrightarrow}}}
 \newcommand\SZ {\emm{{\cal N}}}
 \newcommand\myone {{\mbox{\bf 01}}} \newcommand\mytwo {{\mbox{{\bf 10}}}}
 \newcommand\mythree {\mbox{{\bf 11}}}
 \newcommand\rht{{\emm{\Rightarrow}}} \newcommand\lft{{\emm{\Leftarrow}}}

\title {An \ttl\thanks
 {An abstract of a 20-color version of this result appears in \cite {VL}.
   \newline\hspace*{1pc} This work was supported by NSF grant CCF-1049505.}}
 \author {Leonid A.~Levin\\ Boston University\thanks
 {Department of Computer Science, 111 Cummington Mall,
 Boston MA 02215. \hreff {www.cs.bu.edu/fac/lnd/}}
 \and Ramarathnam Venkatesan\\ Microsoft Research\thanks
 {One Microsoft Way, Redmond WA 98052, venkie@microsoft.com}}
 \date{}\maketitle

\begin {abstract} NP-complete problems should be hard on some instances but
those may be extremely rare. On generic instances many such problems,
especially related to random graphs, have been proven easy. We show the
intractability of {\em random} instances of a graph coloring problem: this
graph problem is hard on average unless all NP problem under all samplable
(i.e., generatable in polynomial time) distributions are easy. Worst case
reductions use special gadgets and typically map instances into a negligible
fraction of possible outputs. Ours must output nearly random graphs and avoid
any super-polynomial distortion of probabilities. This poses significant
technical difficulty.
 \end {abstract}

\section {Introduction}

Many NP-complete problems are easy for random inputs: see, e.g., \cite {Kr76}.
Reductions $A\le_f B$ between NP problems preserve only the worst case
hardness: some instances of $B\supset f(A)$ are at least as hard as those of
$A$. \cite {Jo84}, \cite {Gu} give a good account of some of the issues in
average case analysis of NP-problems. For instance, the Hamilton path problem
takes linear average time: \cite {a}, \cite {GS87}. In such cases,
NP-completeness may be misleading as an evidence of hardness, say for
cryptography. (Many Crypto applications are based on pseudorandomness,
pioneered in \cite {Blum82}, \cite {GGM}, \cite {Yao82}, those in turn require
hard on average one-way functions.) Such a disappointment has happened with the
Knapsack problem: \cite {Sh82}, \cite {La83}.

Another misinterpretation may cause premature abandoning of the search for
algorithms efficient on all but extremely rare and peculiar instances. Some
such algorithms are neat and simple, e.g., the graph isomorphism algorithm in
\cite {EBS80}. Some problems (decoding of linear codes, versions of graph
coloring and independent sets, etc.), however, have eluded such ``average case
attacks''. So, one needs stronger hardness results related to ``typical'' or
``average'' instances of the problem. Such hardness would be sensitive to the
choice of a particular NP-complete problem and its input distribution.

As an evidence that a problem with a particular distribution is ``hard on
average'', we use a notion analogous to NP-completeness: our problem has no
fast on average solution, {\em unless every NP problem under every samplable
distribution has one}. In that case there is no chance to generate, in
polynomial time, hard NP instances and the P=?NP question becomes academic.
Other examples (Tiling, Matrix Decomposition, etc.) are given in \cite {Le86},
\cite {G90}. These papers show that some NP problems with uniform probability
distributions are {\em average case complete}. This concept was further
analyzed in \cite {BCGL}, \cite {il}, \cite {VR}, \cite {Gu95}, \cite {JW},
\cite {aj}, \cite {Le03}, \cite {mr}, and other works.

In typical NP-completeness proofs $A\le_f B$, the reduction $f$ of $A$ to $B$
uses special structures (gadgets) occurring only in a negligible fraction of
instances of $B$ except under strange distributions such as the one generated
by $f$ itself. Thus, the ``hard instances'' of $B$, may be extremely rare, and
the problem may be easy on average under the distributions of interest. The
need to avoid concentrating outputs to negligible sets presents a significant
difficulty in designing $f$. Our $f$ must output uniformly (within a polynomial
factor) random graphs and only use gadgets available in them.

Below, we show the first such simple problem on graphs. This hardness result is
to be contrasted with previous works that used random graphs only to point out
that many NP-complete problems are easy on average. Moreover, as \cite {Gu87}
has shown, average case completeness of a {\bf random graph} problem is {\bf
unlikely} (unless DEXP=NEXP) without {\bf introducing randomizing reductions as
we do here}.

To motivate our problem, we restate the classical edge-coloring problem (where
all edges incident on a node must have distinct colors) in terms of \trm
{3-graphs}, i.e., 3-node induced subgraphs with induced colors and nodes
relabeled 1,2,3: Given a simple graph and $m$ colors, edge-color it so that no
3-graph contains two edges of the same color. We generalize this notion by
allowing the list $C'$ of permissible types of 3-graphs to be arbitrary. As
$C'$ poses only local restrictions, we specify an additional global parameter:
the number $l$ of edges to be left blank. Set $C=(C',l)$. This formalism has
notable power to express many graph problems, even using only one color and
blank. One example is the matching problem on a $2l$ node graph: $C'$ contains
3-graphs with at most one blank edge each. Another is a similarly restated
problem of finding $l$-node clique in given graph: $C'$ restricts blanks to be
self-loops, all pairwise connected. However, our results below have no direct
bearing on the complexity of these classical problems on random instances.

Our graphs are directed (digraphs). We will color some of their edges and loops
with 3 colors, leaving $l$ edges blank. So, each of the 9 edges of a 3-graph
has 5 options: to be colored red, green, yellow, left blank, or be absent. $C$
is taken by picking at random a number $l\in[1,n^2]$ and a subset $C'$ of all
$5^9$ possible 3-graphs. A $C$-colored graph is a graph so edge-colored that it
has $l$ blank edges and all its 3-graphs are in $C'$. In the above examples a
random $C'$ is correct with a constant probability (if $m=O(1)$); $l$ is with
probability $1/n^2$.

Now, our problem admits a simple statement: Given a random digraph $G$ and a
randomly chosen $C$, $C$-color $G$. Using a randomized reduction we show the
problem is complete in the average case; thus, if this problem turns out easy
on average, complexity-based cryptography will be impossible. Without requiring
$l$ blanks, the problem is trivial, but even with only one color, say red, (and
blanks) no polynomial time algorithm is known: $C$ can restrict all blank edges
to be self-loops on a tournament.

We now compare our completeness with other proposed notions of average case
intractability: they reduce the worse case instances of a problem $A$ reputed
to be hard to random instances of a problem $B$. Some such $A$ are
self-reducible (i.e., $A=B$). Or $A$ is a variant of an NP-hard problem $A'$
with altered parameters.
 \\
 In both cases $A$ need not be NP-hard. We give some examples. Solving a
 noticeable fraction of instances would allow solving {\em all} instances for
 random self-reducible problems such as taking square roots modulo a composite,
 or discrete logarithm\footnote
 {over $GF(p)$ (find $x$ from $(p,g{\in}\Z^*_p, g^x)$) or one elliptic
  curve or all (\cite{jmv09}) elliptic curves of the same order $\bmod\,p$.
  The best known algorithms (see \cite {LL86}) run in
  conjectured time $exp((\|p\|(\log\|p\|)^2)^{1/3})$.} These examples randomize
only arguments but not the modulus. \cite {aj} randomizes the entire instance
for a problem: Given $k,q$ and a random matrix $A$, solve $Ax{\equiv}0\pmod
q,\,\|x\|{=}k$. This is shown to be at least as hard as the worst case versions
of finding, up to a polynomial approximation factor, a Shortest Vector or a
Closest Vector or a basis with the smallest orthogonality defect in lattices.
The approximation factor is what separates these problem from NP-hardness
making such reductions (and analysis of lattice based cryptosystems)
possible.\footnote
 {See, e.g., \cite {AR}. Analogous rounding problems in non-abelian discrete
 groups have tighter P-time approximation limits: any factor depending
 only on ambient dimensions would imply P=NP; see \cite {bmv15}.} The above
examples may turn out to be easy on average while some similar problems would
still be hard. In contrast, our completeness assures average hardness, unless
no hard on average problems exist. The above examples have extra structure,
attractive to cryptographic applications. Thus, if hard, they provide {\em
one-way functions}, a very desirable tool similar to hard on average problems
but not known to follow from the existence of the latter. And even from
existence of one-way function, the hardness of the mentioned problems does not
follow.

\section {Basic Concepts and Claims.}

It is essential to consider reductions to (or hardness of) problems with {\em
specific} distributions. If $A$ is hard on average under some distribution, and
$A\le_f B$, then so is $B$ under the induced distribution of $f$'s outputs. But
the latter may not correspond to any distribution of interest. It is
interesting, to show the reducibility under some common distributions, e.g.,
uniform distribution on graphs. To preserve their average case complexity, our
reductions must map typical instances of one problem into typical instances of
another. The definitions of distributions and averaging run times involve
subtleties, some discussed in this section.

Our strings and logarithms are binary; $\St{=}\{0,1\}^*$; $\|x\|$ is the
bit-length of $x$; $(x,y)$ is a string representing a pair of strings $x,y$.
Algorithms' average run times may be considered on inputs with probability
distributions on $\{0,1\}^n$, e.g., the family $\l_n(\{x\})=2^{-n}$. This may
be artificial sometimes: the code lengths of two strings may be the same in
some binary encodings of instances but differ in others. Instead, distributions
may be over the entire $\St$, such as the uniform one $\l(\{x\})= \frac
{2^{-n}} {n(n+1)}$. But for tighter analysis on specific lengths, distributions
$\mu_n$ may be conditioned to $\{0,1\}^n$. We say a property holds for \trm
{almost every (a.e.)} $x$ if it fails with probability $\mu_n=o(1)$. We may
also consider uniform distribution on $\{0,1\}^\N$ and use a \trm {cut-off}
algorithm $C:\{0,1\}^\N\rightarrow\St$ selecting a finite prefix of a stream of
bits after rejecting all shorter prefixes.

These probabilities differ by $\|x\|$ factors, but our completeness results
ignore polynomial factors in probability and, thus, in average running time.
This is similar to the worst case reduction theories, which typically ignore
polynomial factors in run times. So, we use $\l$ for any of these types of
uniform distributions when unambiguous. Our distributions may sum to ${<}1$
allowing processes a chance to fail in producing outputs. $\bar R(\mu)$ is the
output distribution of $R$ on $\mu$-distributed inputs; $\bar R(\l)$ is \trm
{samplable} if $R$ runs in polynomial time \trm {(P-time)} with respect to
(\trm {w.r.t.}) output lengths. We use standard asymptotic notations
$\Theta,O,o,\sim$.

\subsection {Random Inversion Problems and Reductions.}\label {sec-reduce}

Our version of NP is constructive: the witnesses must be explicitly produced.
It is convenient to refer to them in the inversion form: for an algorithm $f$
with $\|w\|=\|f(w)\|^{O(1)}$, given an \trm {instance} $x$, find a \trm
{witness} $w\in f^{-1}(x)$. E.g., $f((a,b))$ may output the graph $a$, if $b$
is its Hamiltonian cycle (or else an error). A \trm {random inversion
NP-problem (RIP)} is a pair $(\mu,f)$ where $f$ is computable in P-time and the
distribution $\mu$ of the instances is samplable. The choice of a particular
distribution $\mu$ is significant: different distributions may concentrate on
different subsets of instances thus specifying completely unrelated problems.
RIPs require actually finding witnesses for instances, so our reductions are
pairs $(R,Q)$ of algorithms to map~both.
 \\ \vspace{6pt}\noindent
 $f_\mu$ is $f(\St)$ excluding $x$ with $\mu(\{x\}){=}0$.
 $R$ is \trm{$(\mu,f)$-injective} if $R(x){\ne}R(y)$
 for all $x\in f_\mu$, $y\in f(\St){\setminus}\{x\}$.

\BD\label {d-red} A \trm {reduction} of a RIP $(\mu,f)$
 to $(\nu,g)$ is a pair $(R,Q)$ of P-time\footnote
 {The P-time requirement is essential for $R$ only on average over $x$.
 This allows a possibility to generalize completeness to problems,
 whose worst-case versions are not NP-complete. Also $\mu$
 may be not a samplable distribution, only dominated by one.}
 algorithms such that:
 \vspace{-4pt}
 \begin{enumerate}
 \itemsep-2pt\parsep-2pt
 \item\label{red-R}
 \makebox[16pc][l]{
 $R$ is $(\mu,f)$-injective and $R(f_\mu)\subset g(\St)$,
 }
 i.e., $R$ maps solvable instances to distinct solvable ones;
   \item\label {red-Q}
 \makebox[16pc][l]{
 $R(f(Q(w))){=}g(w)$,
 }
 i.e., $Q$ solves $f$-instances $x$ given a $g$-witness for $R(x)$;
   \item\label {red-mu}
 \makebox[16pc][l]{
 $\bar R(\mu)(\{y\})\le\nu(\{y\})\|y\|^{O(1)}$,
 }
 i.e., $R$ maps $\mu$-typical instances into $\nu$-typical ones.
   \end{enumerate}\ED

The reductions are closed under composition and any fast on average algorithm
for $(\nu,g)$ yields a fast on average algorithm for $(\mu,f)$. We consider two
generalizations. A \trm {padded} RIP $(\nu,g)$ is a modification of a RIP
$(\nu,g')$ where $g((w,\a))=(g'(w),\a)$. The distribution $\nu$ is only on
$y{=}g'(w)$, combining all $\a$. (Padding $(\nu,g)$ above with $R$'s input
circumvents the injection requirement.) A \trm {randomization} $(\mu,f)$ of
$(\mu',f')$, is a RIP where the padding $\a$ is also random. It includes a
``cut-off'' P-time algorithm $C(x,\a)$ that selects the padding length. The
digits of $\a$ are chosen at random but $\a$ is restricted to sets $s_x$ of
measure $\l(s_x)=\|x\|^{-O(1)}$ (in our case actually $\sim1$). Then
$\mu(\{(x,\a)\})$ is $\mu'(\{x\})/2^{\|\a\|}$ for $\a\in s_x$, or $0$ for
$\a\notin s_x$.

\BR $s_x$ consists of those ``helpful'' strings $\a$ which allow $R$ to output
instances $y$ whose witnesses can be transformed by $Q$ into witnesses for $x$,
if any. $s_x$ is not required to be decidable in P-time, so this artificial
restriction breaks samplability of $\mu$. But $s_x$ has a polynomial
probability, so a randomized algorithm can try several $\a$, shrinking
exponentially the probability of failure for solvable instances. Some
algorithms inverting $g$ give also negative answers ``no inverses exist.''
Unlike inverses, they are not verifiable and so can be used in reductions only
if their chance exceeds by $\|x\|^{-O)1)}$ a known $\l(s_x)$ bound.\ER

\BD \label {d-compl} A RIP $(\nu,g)$ is \trm {complete} if every RIP has a
randomization reducible to it (or to its padding).\ED

\pp {The \trm {average complexity}} of $(\mu, f)$ is an upper bound $T(k)$ for
the time needed to find $w{\in}f^{-1}(x)$ expressed in terms of $k(x){=}\|x\|r
(x)$, where the instance's ``rareness'' or ``exceptionality'' $r$ has $\sum\mu
(\{x\})r(x)<\infty$. Reductions and randomizations preserve average complexity
(up to a polynomial) and completeness. Note that complete problems {\em must}
have many easy instances, since both hard and easy problems are reducible to
them. Note also that if two distributions $\mu$ and $\mu'$ are such that $\mu
(\{x\})/\mu'(\{x\}) =\|x\|^{O(1)}$, then $R(x)=x$ reduces $(\mu,f)$ to $(\mu',
f)$: the polynomial factors in distributions are absorbed by the definition.

\subsection {Main Result} \label {sec-mr}

Let $\cal G$ be a digraph with edges (including, \trm {loops}, i.e.,
self-loops) colored red, green, or yellow, or left blank. Blank edges play a
different role than colored ones in that we limit the number of colored
(non-blank) edges.

A \trm {spot} in $\cal G$ is a 3-node subgraph with induced edges and colors,
and the nodes unlabeled (i.e., replaced by 1, 2, 3). So there are $<5^9$
distinct spots. The \trm {coloration} $C({\cal G})$ is the set $C'$ of all
spots in $\cal G$ and the number $l$ of its blank edges; $G{=}\ov {\cal G}$ is
obtained from $\cal G$ by removing the colors. As noted in the introduction,
the classical edge coloring problem restricts the set of spots so that all
edges in a spot have different colors. We choose a random restriction instead.
We similarly have 2-node spots called \trm {links}.

\pp {Graph Coloration Problem:}
 \ \\
 Invert the function $f({\cal G})=(\ov{\cal G},C({\cal G}))$, i.e.,
 color the edges of a given graph to achieve the given coloration.
 {\bf Uniform distribution:} $\l(\{G\})=\frac{2^{-n^2}}{n(n+1)}$, $\l(\{C'\})
  {=}O(1)$, $\l(\{l\})=n^{-2}$ for all $C',l$, and $n$-node graphs $G$.

\BT Graph Coloration is a complete random inversion problem. \label {main}\ET

We could use as well a monotone variation, accepting also any smaller $C'$
and/or larger $l$. The function to invert would then map $({\cal G},C',l)$ to
$(\ov{\cal G},C',l)$, if $C({\cal G})=(\ov {C'},\ov l)$, $\ov{C'}\subset C'$,
$\ov l\ge l$, to error otherwise.

\subsection {Outline of the Proof}

The rest of the paper is devoted to proving this theorem. Section~\ref
{sec-gnp} has several lemmas on random graphs to show they are likely to have
the structures (gadgets) needed for our reductions. (Some of these properties
require tighter bounds than the literature offers so we must go into some
computations.) One is the likely existence of a unique $(\log n)$-node
tournament $T$. Edges between a node and $T$ define bits of a string we call
node's \trm {code}, used to order some nodes. Another property to mention is a
bound on probability of existence of matchings that are used to embed specially
designed large graphs of bouded degree into random graphs.

Section~\ref {TM} defines a restricted type of Tiling Problem RRTP which
retains its completeness and which we reduce to our graph coloring. Limited to
3-color 3-node patterns, we must carefully select a universal Turing machine,
split its symbols into bits/trits, and economically represent them in the
tiling. We use a UTM from \cite {I58} and utilize its undefined transitions to
implement our extras, e.g., non-deterministic choices.

A tiled square is encoded onto a colored grid (represented as a graph) which we
call a template. The grid squares have graph's looped nodes at the corners and
unlooped nodes in the centers. The edges connecting corners with adjacent
corners and with the centers are to be colored to reflect the tiling symbols.
The reduction is based on embedding this template into a random graph. The grid
nodes are those connected with all tournament nodes. The centers are placed in
an order determined by their codes.

Section~\ref {R} presents our reduction. It generates roughly uniformly random
graphs with an embedded instance of our random tiling problem. We generate a
graph at random, except for several features imposed on it. These features are
likely to exist on their own, but may be hard to discover (e.g., it is not
known how to find a $(\log n)$-node tournament in polynomial time). Some
features are imposed for convenience, to reduce to $o(1)$ the probability of
unsuitable graphs. Among these features are counts of certain types of nodes to
match their expected value, the codes mentioned above being distinct, etc. One
feature is of a different nature: The edges between successive grid centers
must reflect the input bits of the tiling instance. This feature appears in
random graphs only because the instance of Tiling is itself randomly chosen.

Section \ref {sec-grid} describes the required coloring patterns of all 3-node
induced subgraphs and proves the equivalence of solving the Tiling Problem to
the graph coloring it is reduced to. The tournament $T$ is marked with blank
loops; its two-way connections to looped nodes are allowed to include blank
edges. The limit on the non-blank edges then assures the $\log n$ size of $T$,
which renders it unique. We then transform a solved Tiling instance into a
template colored graph and use a lemma from section \ref {sec-gnp} to embed
this graph into the graph produced by the reduction. Any coloring of the graph
to exhibit a solution to the GCP instance is forced to mark a grid on all nodes
connected to the whole tournament. The colors of the grid edges must display
the tiling solution. This used to be the trickiest part of the construction but
is now much simpler.

\section {Random Graph Lemmas} \label{sec-gnp}

We describe the structures needed for the reductions and prove that they exist
on a polynomial fraction of graphs. Such questions, when restricted to
properties of almost all graphs, are addressed in the theory of random graphs
(see \cite {Bo01}). Transitive tournaments (thereafter simply tournaments) are
complete acyclic, except for loops at each node, digraphs. $\E_\mu(Y)$ stands
for the expected value of the random variable $Y$ w.r.t. the distribution
$\mu$. We omit $\mu$ when the distribution is understood. Unless stated
otherwise, our random graphs have edges drawn independently with probability
$1/2$. In this section we consider distributions on graphs with a given (not
chosen at random) number of nodes. $f\sim g$ means $f(n)=(1+o(1))g(n)$.

\BR\label {l-prob} We use the following facts on probability and
 approximations: \begin{enumerate} \item \begin{enumerate}
 \item\label {r21a} $n!=\left(\frac{n}{e}\right)^n \sqrt{2\pi n+
 \theta_n}$, where $\theta_n\in[\pi/3,~e^2{-}2\pi]\subset[1,~1.11]$.
 \item\label {r21b} For $k=o(\sqrt{n})$, $\binom nk\sim{\frac{n^k}{k!}}$,
 \item Let $k=o(\sqrt n)$ integers be chosen from $\{1, \ldots, n\}$
 randomly and independently.
 \\
 The probability that all are distinct is
 $\Pi_{j=1}^{k-1}(1{-}\frac{j}{n})=e^{-\Theta(k^2/n)}\sim 1$. \end{enumerate}

\item Let random variable $Y$ take values $i$ with probability $\Pb(\{i\})=
 p_i>0$. $i\in \{1,\ldots,s\}$.
 \\
 Let $y_i$ be the number of occurrences of
 $i$ in $n$ independent Bernoulli trials of $Y$.
 \\
 Then, {\em from the above
 item~\ref {r21a} and the central limit theorem:}\begin{enumerate}
 \item $np_i{=}\E(y_i)$; $\s^2_i=$ Variance$(y_i)=np_i(1-p_i)$
 \item\label {r22b} If $x^3=O(\s_i)$, then $ {|y_i-np_i|}
 \ge \s_i x$ with probability $e^{-x^2/2}/\Theta(x+1)$
 \item Let $k_i$ be integers with $\sum_i k_i=n$, $t_i$ be
 $\frac{k_i}{np_i}{-}1$, and $\sum_i np_i t_i^2 =O(1)$.
 \\
 Then $\Pb(y_i=k_i,\,i=1,\ldots,s) =\sqrt{n}/O((2\pi n/s + 1.11)^{s/2})$.
 \end{enumerate} \end{enumerate} \ER

\BPR [Proof of item 2(c).]
The number of ways in which the event $[y_i=k_i,i:=1,\ldots,s]$ can occur is
 \\
 $n!/\prod_{i\le s} (k_i!)$
 and each of these has probability $\prod_{i=1}^sp_i^{k_i}$.
 Thus the required probability is \[n!\prod_{i=1}^s\frac{p_i^{k_i}}{k_i!}=
n^n\prod_{i=1}^s\left(\frac{p_i}{k_i}\right)^{\!k_i} \sqrt{\frac{2\pi
n+\theta_n}{\prod_{i{=}1}^s(2\pi k_i+\theta_{k_i})}}\]
 Now, $n^n\prod_{i{=}1}^s \left(\frac{p_i}{k_i}\right)^{\!k_i}=e^{-n(\sum_i
  p_i(1{+}t_i)\log(1{+}t_i))}$, which we will now see is $1/O(1)$.
 Indeed, $\log(1{+}t)\le t$,
 \\
 so $\sum_i np_i(1{+}t_i)\log(1{+}t_i)\le \sum_i np_i t_i+
 \sum_i np_i t_i^2 =0+O(1)$, by the assumptions.\EPR

\BL\label {l.trn} Let $X_k$ be the number of $k$-node tournaments
 in a random $n$-node digraph,
 \\
 $c=k{-}\log n$, $|c|{=}o(k)$, $t=2^{-kc}$.
 Then $\E(X_k)\sim t$; $\E(X_k^2)\sim t+t^2$.\EL

\BPR First, note that $X_k^2=\sum_l N_l$, where $N_l$ is the number of ordered
pairs of $k$-node tournaments sharing $k{-}l$ nodes. By linearity of
expectation, $\E(X_k^2)=\sum_l\E(N_l)$. There are $\binom kl$ ways to choose
the positions of shared nodes in each tournament in a pair of $k$-node
tournaments and $\binom n{k+l} (k+l)!$ ways to map this structure in our graph.
The probability of each pair to form the tournaments is $2^{-(k^2-l^2+2kl)}$.
 \\
 Using item~\ref {r21b} of the Remark~\ref {l-prob} we get
 $\binom n{k+l} (k+l)!\sim n^{k+l}$ and substituting $n=2^{k-c}$ below,
 \[\E(N_l)\sim 2^{l^2-k^2-2kl}\binom kl^2n^{k+l}=\binom kl^2 2^{-c(k+l)}2^{l(l-k)}.\]
 Now, $\E(X_k)=\E(N_0)\sim t$, $\E(N_k)\sim t^2$, and
 $\E(X_k^2)=\sum_{l=0}^{k}\E(N_l)\sim t+t^2+ \sum_{l=1}^{k-1}\E(N_l)$.
  \\
 It remains to show that $\sum_{l=1}^{k-1}\E(N_l)=o(t{+}t^2)$.
 Note that for $0{<}l{<}k$, $2^{-c(k+l)}\le 2^{-|c|}(t{+}t^2)$. So,
 \[\E(N_l)/(t{+}t^2)\le\binom kl^2 2^{l(l-k)} \le
 \begin{array}{ll} 2^{-l((k-l)-2\log k)} &\text {for } 0{<}l{<}k/2\\
 2^{-(k-l)(l-2\log k)} &\text {for } k/2{\le}l{<}k.\end{array}
 \le2k^2/2^k=o(1/k).\]\EPR

\bigskip Let $0<c+1/c=o(k)$ and $\l_c$ be the probability distribution of a
$\floor{2^{k{-}c}}$-node graph $G$ generated as follows: First we choose
randomly a sequence of $k$ nodes of $G$ and force a tournament $T$ on them.
Then, we choose at random the set of all other edges $(i,j)\not\in V(T)^2$. The
following corollary implies that a random graph on $\floor{ 2^{k{-}c}}$ nodes
has a unique $k$-node tournament with a polynomial probability if and only if
$c=O(1)$.

 \BC\begin{enumerate}
 \itemsep-2pt\parsep-2pt
 \item For those $G$ with unique $k$-node tournaments,
 $\l_c(\{G\})/\l(\{G\})\sim 2^{kc}$.
 \item $\l_c$-almost every $G$ has only one $k$-node tournament.
 \end{enumerate}\label{c.max}\EC
 \vspace{-9pt}

\BPR First note that $\l_c$-probability of all such pairs $(T,G)$ is the same,
and the total number of such pairs is $\sum_GX_k(G)=\E_\l(X_k(G))/\l(G)$. Thus,
$\l_c(\{G\})/\l(\{G\})=X_k/\E_\l(X_k)$, which is $\sim2^{kc}$ if $X_k{=}1$.
Then, the expected number of tournaments, other than the forced one, is $o(1)$:
$\E_{\l_c}(X_k)=\sum_GX_k(G)\l_c(\{G\})=\sum_G X_k(G)\frac{X_k(G)\l(\{G\})}
{\E_\l(X_k)}=\frac{\E_\l(X_k^2)} {\E_\l(X_k)}\sim
\frac{2^{-kc}+2^{-2kc}}{2^{-kc}}=1+2^{-kc}\sim 1.$\EPR

We use the next lemma on matchings in random graphs to show that a.e.\@ $G$
contains any $O(1)$-degree graph (e.g., Hamilton path, grids) as a spanning
subgraph. Lemma~7.12 of \cite {Bo01} proves a.e.\@ graph has perfect matchings.
We need a tighter bound so use tighter calculations and a ``no solitary edge''
argument:

\BL\label{match} Let $G$ be a bipartite undirected graph with edges in
$V_0\times V_1$ chosen independently with probability $b/n$, $n=|V_i|$. Then,
for large enough $n$, the probability $\s$ that $G$ has no perfect matching is
$<2n/e^b$.\EL

\BPR By Hall's Theorem, if $G$ has no perfect matching, it has an
$(n{+}1)$-node independent set $L_0{\cup}L_1$, $L_i{\subset}V_i$. Let $L$ be a
smallest such $L_i$ in $G$. Then $a\,\edf\,|L|{-}1<n/2$, and no node has
exactly one edge to $L$ (otherwise it can replace its neighbor, shrinking $L$).
Let $\s_a$ be the probability of $G$ containing such an $L{=}L_0$. Then $\s<2
\sum_{a=0}^{(n{-}1)/2}\s_a$. Assume that $b{<}n{<}e^b/2$, otherwise Lemma is
trivial. A node is isolated with probability $p=(1{-}b/n)^n <e^{-b-b^2/2n}<
e^{-b}/(1{+}b^2/2n)$. $L_0,L_1$ and two edges from each $x\in
V_1{\setminus}L_1$ to $L_0$ have $C_a=\binom n{a{+}1}\binom na\binom{a{+}1}2^a$
choices. For $a\in[2,\frac n2]$, $h\,\edf\,
1{-}\frac1a{-}\frac{a+1}n\ge\frac12{-}\frac 3n$, using $\binom
nk{<}\left(\frac{ne}k\right)^{\!k}$, we have:
 \[
 \s_a<C_a\!\left(\frac bn\right)^{\!2a}\!\!\left(1{-}\frac bn
 \right)^{(n-a)(a{+}1)}\!\!<\!\left(\frac{ne}{a{+}1}\right)^{\!a{+}1}\!\!\!
 \left(\frac{ne}a\right)^{\!a}\!\left(\frac{a(a{+}1)}2\right)^{\!a}\!\!
 \left(\frac bn\right)^{\!2a}\!\!\!p^{ha{+}2}
 =\left(\frac{e^2b^2p^h}2\right)^{\!a}\!\!\frac{p^2ne}{(a{+}1)}=o(p^2n).
 \]
 By inclusion-exclusion, $\s_0{<}np{-}\binom n2p^2{+}\binom n3p^3 <
 np(1{-}p\frac n3)$. Also, $\s_1<\binom n2n(\frac bn)^2p^{2{-}2/n} <
 \frac n2b^2p^{2-2/n}$. So, $\s<2\s_0{+}2\s_1{+}o(p^2n)n<
 2ne^{-b}(1{-}\frac{pn}3{+}\frac{b^2}2p^{1{-}2/n}{+}o(pn))/(1{+}\frac{b^2}{2n}){<}
 2ne^{-b}.$ Indeed, $\frac{{-}pn}3+\frac{b^2}2p^{1{-}2/n}{+}o(pn)<b^2/2n$,
 or equivalently, $\frac bn+\frac nb(2/3{-}o(1))p >p^{1-2/n}b$ by using
 $x{+}y\ge2\sqrt{xy}$ and $b<|\log p|$.\EPR

 \BL [Equitable Colouring: \cite {HS}\label {lem-part}]
 Any undirected graph of degree $<d$
 \\
 can be partitioned into $d$ independent sets
 whose sizes differ at most by one. \EL

See \cite {Bo04} for a proof. \cite {KK} gives a P-time algorithm but we need
here only the existence of the partition. Next we use this lemma to find
embeddings in random digraphs of some structures we need in our reduction. We
first describe these embeddings.

Write $H \sqsubset G$ if $H$ is an induced subgraph of $G$. We call nodes $u$
and $v$ \trm {connected by a single edge} if exactly one of the two directed
edges $(u,v), (v,u)$ exists, by a \trm {double edge} if both do, and \trm
{non-adjacent} or \trm {disconnected} if neither does. The degree $\dgr(u)$ of
a node $u$ is the number of adjacent nodes, and $\dgr(G)$ is the maximum degree
of nodes in $G$. A \trm {di-embedding} $g:V(G)\hookrightarrow V(H)$ is a
bijection isomorphic on each {\em connected} 2-node subgraph of $G$. Note,
disconnected nodes can be mapped to nodes that may be connected. We call two
graphs \trm {compatible} compatible if they have the same number of looped and
unlooped nodes.

\BL [Embedding Lemma\label{l.emb}] Let $H{\sqsubset}F$, $d:=\dgr(F)$, and
$m:=|V(F\setminus H)|=4^dd^3/o(1)$.
 \\
 Choose digraphs $G$ by dropping all non-loop edges incident on $F\setminus H$
and putting randomly chosen
 \\
 edges instead. For a.e.\@ such graph, there exists a di-embedding $g$ of $F$
into $G$ which is an identity on $H$.\EL

We will use this lemma for a graph with $d=O(1)$: a grid with some additional
edges. Now we want to find this grid in the random digraph by constructing a
suitable $g$. We reduce the task of constructing $g$ to that of finding perfect
matchings in a sequence of random bipartite undirected graphs: using the
lemma~\ref {lem-part} we partition $F\setminus H$ suitably into some
independent sets $I_j$ and extend $g$ on each $I_j$ one by one.

\BPR {\bf Initialization:} First, we partition $F\setminus H$ into independent
sets $I_j, j{\le}d^2{+}1$, such that $ |I_j|\ge m':=\floor{m/d^2}$, and each
$I_j$ consists of either only looped nodes or only unlooped nodes. Let $F'$ be
a graph obtained by adding edges between nodes in $F\setminus H$ that share a
neighbor in $H$. Clearly $\dgr(F')< d^2$. Now, $F'$ can be partitioned into
independent sets using the lemma~\ref {lem-part} as needed. Next, we split
$V(G\setminus H)$ into disjoint $V_j$, compatible with $|I_j|$. (The latter
condition can be met since $G,F$ are compatible.) Set $K=H$. Below $K$ will
denote the set of nodes on which $g$ is defined (i.e., di-embedding is found).

{\bf Repeat} the following steps for $j:=1,2,\ldots$, until the extension of
$g$ is completed.

{\bf Extension Step(j):} Call $v{\in}V_j$ a \trm {candidate} for $u{\in}I_j$ if
setting $g(u):{=}v$ will extend $g$ as a di-embedding over $K\cup\{u\}$. For
this, $v$ must satisfy at most $2d$ connectivity constraints specifying which
of the directed edges are present. Now we construct an undirected bipartite
graph $G'_j$ by connecting every node in $I_j$ to all of its candidates in
$V_j$. Then, with a high probability (over $G$), we find a perfect matching
that maps $I_j$ to $V_j$ satisfying the connectivity requirements. This way we
extend $g$ over $K\cup I_j$ and update $K$ to $K\cup I_j$.

{\bf Analysis of Extension Step (j):} In $G'_j$ the probability of an edge
$\{u,v\}$ is $4^{-\dgr(u)}$. Also the edges are chosen independently: the
conditional probability of $v$ being a candidate for $u$ is the same, given the
set of all the edges from $v$ to $\Gamma(I_j-\{u\})\cap K$, since
$\Gamma(I_j-\{u\})$ and $\Gamma(\{u\})$ have no nodes in common in $G_j'\cup
K$, which is guaranteed by construction of $F'$ in the initialization and
partitioning steps. The existence of matching is a monotone property i.e.,
cannot be broken by adding edges. So, it suffices to estimate its probability
by decreasing the chance of all edges in $G_j'$ to the uniform $4^{-d}$. Now we
come to the overall success probability (over $G$) of finding $g$. Put
$b/m'=4^{-d}$. Then, using lemma~\ref {match}, the probability that one of the
$d^2$ matching steps fails is $O(d^2 m/d^2)e^{-b}=O(m)/e^{m'/4^d}
=O(e^{-(\frac{m}{d^2 4^d}-\log m)})=o(1).$\EPR

\section {Turing Machines and Tilings.}\label{TM}

Our proof uses the completeness of the \trm {Random Tiling Problem (RTP)}. By
\cite {il}, any samplable Random Inversion Problem (RIP) reduces to a problem
with P-time computable distribution, i.e.\@ with the measure of intervals $[0,
x]\subset\N$ computable in time $\|x\|^{O(1)}$. \cite {Le86} proved RTP to be
complete for such problems using deterministic reductions. A tile is a square
with each corner labeled by a letter A-Z. A tiling of an $n\times n$ square,
involves covering it with $n^2$ tiles, so that the letters on the touching
corners of adjacent tiles match. Our RIP will be inverting a function $\cal T$
that, given a tiled square input, returns the set of tiles (called \trm
{legal}) used in it, and its lowest (\trm {floor}) row of tiles.

{\bf Problem:} Invert $\cal T$, i.e., given a set of legal tiles and a floor
row, extend it into a tiled square.

{\bf Distribution:} Uniform: choose randomly $n$ with probability
$\frac1{n(n+1)}$, the set of legal tiles from all possible $26^4$ tiles, and a
legal tile at the lower left corner; choose each successive tile of the floor
row with equal probability from the legal tiles matching the previously chosen.
(If none exists, output a trivial instance.)

Our graphs are edge-colored with only 3 colors, so we need careful restrictions
on letters in the complete RTP we use. Figure~\ref{move} below pictures our RTP
representing Turing Machines (TM). Our symbols have four fields: $\trit\in
\{0,1,*\}$ (called a trit) and three bits $b,s, s^-$. We refer to $s,s^-$ as
the \trm {direction} and \trm {previous} direction (\trm {leftward: \lft} or
\trm {rightward: \rht}), to $b$ as \trm {priming} (and picture $b{=}1$ by
priming the trit (e.g., $*\mapsto*'$). No legal tile $(^{x\,y}_{z\,u})$ has
$z_s\ne x_{s^-}$ or $(x_s\,y_s)=(\lft\,\rht)$. Thus each row will consist of
two segments: rightward pointing segment at the left and leftward pointing one
at the right. Tiles where $z_s$ is $\lft$ have $y=u$. If $u_s$ is $\rht$ then
$x=z$. Two tiles cannot agree on both $z,u$ but differ on both $x,y$. The left
segment of the floor row must have $b{=}1$, the right segment -- $b{=}0,
\,\trit{\ne}*$. The top and floor symbols carry a special flag and must agree
so that the entire tiled square can be wrapped into a cylinder. The cylinder
has constant symbols at each end; it is further wrapped into a torus by a ring
of the special \trm {wall} tiles made by repeating this pair of symbols.

 We use just one particular tile set and it meets the above restrictions. Any
tile set has a constant probability among all tile sets, so the density
requirements of the reductions are not affected. Note that tiling an $n\times
n$ square is easily reduced to tiling a larger $2p\times p$ rectangle with a
prime $p\in[n,2n]$. Our argument will use only such rectangles. We reduce such
restricted form of RTP (called \trm {RRTP}) to Graph Coloration Problem. Thus,
we need to see that RRTP restrictions are compatible with reductions of \cite
{Le86} of non-deterministic computations by a universal TM to tiling.

\subsection {Completeness of the RRTP.}

Any RIP can be stated as accepting random instances by a run of a given
non-deterministic P-time TM. All TMs can be simulated by a universal TM with a
polynomial time overhead and some special short (logarithmic) input prefix. We
will now describe such a universal TM of \cite {I58}.

\pp {Turing Machines} work on a tape consisting of a sequence of cells. Exactly
one cell contains the TM's \trm {head state $S$}, the others -- its tape
symbols $\s$. A cell content at a given step is called an \trm {event}. Their
space-time table (i.e., a rectangular grid where $i$-th row contains the entire
tape contents at $i$-th step) describes the history of the computation. The
TM's head follows some path from its bottom row to the top. At each step the
head acts on one of its two adjacent cells. We call this cell, the head's cell,
and the border between them \trm {active}, and others---\trm {idle}. We now
describe some {\em local} rules ensuring the global picture. Cells' content
specifies the direction $s$ to the active border. Thus, adjacent cells cannot
both have $s$-directions facing away from each other. Idle cells do not change
content from one row to the next. Active cells do, performing a \trm
{transition}. Exactly one of them changes direction and that one carries the
head at the next step. Figure~\ref {move} pictures the space-time history
around the active cells for the TM transition moving the head to the right.
 Its box containing the state $S_1$ points toward $\s_1$ (left or right as
reflected in cases (a) and (b) respectively). In case (a), the right move
merely causes the head to flip its direction; otherwise, the head switches
places with the other active cell of the current row. Left moves work
 similarly as per the above rules.

 \newcommand\captL
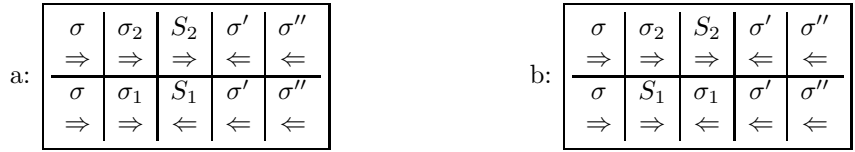
 {Space-Time History for a Right Move $(\s_1,S_1)\to (\s_2,(S_2,RIGHT))$}
 {\begin{figure}
 [hbt]
 \hfill a: \fbox{\(\begin{array} {c|c|c|c|c}\s&\s_2&S_2&\s'&\s''\\
 \rht&\rht&\rht&\lft&\lft
 \\\hline\s&\s_1&S_1&\s'&\s''\\
 \rht&\rht&\lft&\lft&\lft
 \end{array}\)}\hfill
 b: \fbox{\(\begin{array} {c|c|c|c|c} \s&\s_2&S_2&\s'&\s''\\
 \rht&\rht&\rht&\lft&\lft
 \\\hline\s&S_1&\s_1&\s'&\s''\\
 \rht&\rht&\lft&\lft&\lft
 \end{array}\)}\hfill\null
 \caption{\captL}\label{move}\end{figure}}

A binary TM has $\{0,1\}$ tape alphabet. To convert a regular TM into a binary
one $M$ we must deal with its lack of a blank symbol that usually denotes the
end of input. For this we prefix its input $x$ with padding that encodes input
length $l=\|x\|$ as a binary string preceded by a string of $2\|l\|$ zeros. In
the cells carrying this padding, two counters are initiated that monitor the
distance to both ends of the used part of $M$'s tape (initially the input).
$M$'s head moving on the tape pulls these counters along and keeps them
updated. When the right end of the used tape is reached, any subsequent
characters are treated as blanks. $M$ and its counter are initialized so that
the head never approaches the tape ends. The simulated TM $M$ may have a
``write $0$ or $1$'' nondeterministic command. It is used to guess the solution
of the instance. Our RRTP needs the space time history to wrap around making
the top row consistent with the bottom row being next to it. The
non-determinstic command is used for this after a successful end of the
computation.

We describe below a version of a TM $U$ of \cite {I58} that simulates any such
binary $M$. $U$ has $6$ leftward (i.e., with our $s{=}\,\lft$) and $5$
rightward head states and $6$ tape symbols. We represent them by the same
fields $\trit,b$ (plus the directions $s,s^-$) as used in RRTP. Head states
differ from tape symbol by $s\ne s^-$, i.e., changed from its previous value.
The tape of $U$ simulating $M$ consists of a program segment $P$ followed by
the tape of $M$ segment denoted $T$. At each step, $U$ marks the current cell
it is working on, goes all the way left to program segment, decides on the next
step to perform, and returns to the current cell to do so. The tape trits in
segment $P$ never change. In segment $T$, the trits represent the current tape
bits of $M$ except that the $M$'s active tape symbol may be replaced by a $*$.
This $*$ and the priming bits $b$ are the only non-$M$ data $U$ uses. The
simulation starts with $P$ having only $0',1',*'$ symbols followed by the head,
then by $T\in\St$.

The transition table below has the $(*',e)$ command (one of the ``halt''
commands in \cite {I58} denoted by $=$) modified to a choice of entering state
$A$ or $B$. This simulates $M$'s non-deterministic command. We modify the
initial state to be $(*',e)$, too; the effect of the non-determinism of $U$'s
first transition to $A/B$ disappears in 3 steps: easy to see by tracing them.
Afterwards, the simulation proceeds in a regular way.

 {\newcommand\ikntext{
 The table's columns are indexed by tape symbols $\s$, the rows by the states
 (rightward directed -- by uppercase letters, leftward -- by lowercase).
 The entries reflect the resulting state (blank if unchanged), the trit of the
 new $\s$ (blank if unchanged), and its bit (always, even if unchanged, e.g.,
 in the transition from $(A,1')$ the symbol $1'$ gets unprimed, while from
 $(a,1')$ it stays primed).\par The $(D,0)$ command is unused. We employ it
 to mark a column of ``wall tiles'' merging the left border of the space time
 history with its right border to obtain a cylinder. $U$ comes to its
 ``choose $A$ or $B$ state'' command $(*',e)$ if the digit that $M$ is to
 write in the Ikeno's representation of $M$'s command is replaced with a $*$.
 Then $U$ has a choice to act as if it was $0$ or $1$.}

 \noindent\makebox[\hsize]{\parbox
 {285pt}
 {\ikntext}\hfill\fbox{$\begin{array} {c||c|c|c|c|c|c}
    & 1' & 0' & *' & 1 & 0 & * \\ \hline
 A & & & & f & f & e0\\\hline
 B & & & & F & F & e1\\\hline\!\!
 f,F\!\! && & c & b* & a*& F \\\hline
 c & = & F & E\,'& ' & ' & \\\hline
 a & b\,'& F & E\,'& ' & ' & ' \\\hline
 b & & a\,'& D & ' & ' & ' \\\hline
 d & ' & ' & D & ' & ' & \\\hline
 D & & & e\,'& d\,'& - & \\\hline
 E & ' & ' & e\,'& = & - & ' \\\hline
 e & B & A &\!\!A/B\!\!&' &'&'\\\end{array}$}\hfill}}

\section {The Reduction Algorithm R}\label{R}

The reduction $R$ described below produces a graph $G$ from any RRTP instance
$X$ supplemented with
 \\
 a random $O(\|X\|^7)$-bit $\a$. $R$ pairs $G$ with a coloration:
 {\em standard} $C'$ and $l$ described in the next section.
 \\
 We show that $C$-colorings of such $G$ exist for any solvable $X$,
 and all of them easily yield solutions for $X$.

\BR Recall that the reduction (see Sec.\ref{sec-reduce}) involves a pair of
algorithms $R$ and $Q$ such that: \begin{center} (Randomized) tiling instance
$Y=(X,\a)\,\stackrel{R} {\longrightarrow}$ graph coloration instance $(G,C)$
 \\
 Solution $w'$ of tiling $Y$ $\stackrel{Q}{\longleftarrow}\,C$-colored $G$
\end{center} To assure the solvability of $R(Y)$, section~\ref {sec-grid}
transforms a tiling solution to a colored graph template $G'$ and di-embeds
$G'$ into a suitably colored $G$. This can be seen as a (not necessarily
P-time) inverse of $Q$: \begin{center} Tiling solution
$w'\stackrel{Q^{-1}}{\longrightarrow}$ Edge
  colored graph $G'\hookrightarrow G$ \label{rem-three}\end{center}\ER

Now we describe how $R(Y)$ designs the graph $G$.
 \\
 Let a prime $p$ be the grid size of $X$, $k{=}\ceil{\log_{1.5}5p^2}$,
$n{=}\floor {4p^2(4/3)^k}$. First, $R$ randomly selects disjoint sets
$T,U_T,L_T$, in $|G|{=}\{1,\ldots,n\}$ so that $|U_T|=|L_T|{+}1=2p^2$,
$|T|{=}k$. Then $R$ creates a graph $G$ on $|G|=L\cup U$ ($L$ being its set of
looped, $U$ of unlooped nodes) randomly except that \ref{P1}-\ref{P4} are
enforced:

\begin {enumerate} \renewcommand\labelenumi{(\roman{enumi})}
\renewcommand\theenumi\labelenumi\item\label{P1}
 Each node $i$ has $\frac n8{\pm}x_i$, $x_i{<}n^{2/3}$ double edges to $L$. $T$
forms a tournament $t_k,\ldots,t_1$ (source to sink).
 \item\label{P2} $L_T\subset L$, $U_T\subset U$. $U_T\cup L_T\cup T$ is the set
of all nodes connected with every node in $T$.

{\em The next condition uses the concept of \trm {codes}
 defined in \cite {EBS80} as the $2|T|$-bit string, whose $i$-th bit
 reflects the presence of the \trm {code edge} $(u,t_{i/2})$ or, for odd $i$,
 $(t_{(i+1)/2},u)$. The \trm {code} of $u$ \trm {w.r.t.} $t_i{\in}T$
 is the restriction of its code to the digits $2i$ and $2i+1$.}

\item\label{P3} All codes of $U_T$ differ even in their $3/4$
  most significant digits (\trm {$\mathbf{\frac34}$-codes}).
 Let $v_1,v_2,\ldots v_{2p^2}$ be\\
 all the $U_T$ nodes in decreasing order of codes
 and $v_d$ be the first one disconnected from $v_{d+1}$.

\item\label{P4} $t_1$ has double edges to $v_1,v_p,v_{2p^2{+}1{-}p},v_{2p^2}$;
  the four connection types of $v_i$ to $v_{i+1}$, $i\le d$ reflect our RRTP
  floor's left ({\lr} is $\rht$) symbols $\tau$; for $d<i<p$, the forward
  $(v_{i-1},v_i)$ edges reflect its right $\tau$ bits. \end {enumerate}

\pp {Denote}
 \hspace{-6pt}
 $\SZ_2$ the set of graphs with unique $k$-node tournament,
 respecting~\ref{P1},\ref{P2}; $\SZ$ also respect \ref{P3},\ref{P4}.

 \BP\label {p.codes}
 In a.e.\@ $G$ all nodes in $U_T$ have distinct $\frac34$-codes. \EP

\BPR The codes of nodes in $U_T$ are independent and uniformly distributed for
graphs in $\SZ_2$. There are $x=3^{3k/4}$ possible $\frac34$-codes for $U_T$,
and $|U_T|<y={(\frac32)}^k$. The chance that all codes are distinct is
$(1-1/x)^{y(y-1)/2}>e^{-y(y-1)/(2x-2)}>e^{-y^2/x}\sim1$, since
$y^2/x=((3/2)^2/3^{3/4})^k=(243/256)^{k/4}=o(1)$.\EPR

\BR\label{rem.uni} A.e.\@ graph generated by $R$ has a unique $k$-tournament,
unique codes for $U_T$, and so is in $\SZ$ and uniformly distributed on it.
Thus, we will consider the uniform distribution on $\SZ$ and denote it by
$\Pb$. We define $s_X$ in section~\ref {sec-reduce} to contain exactly those
$\a$ in $Y$ resulting in graphs $G\in\SZ$ that satisfy all properties of (a.e.)
$G$ required below. Thus, these properties of $G$ will be assumed.\ER

The theorem follows from the two following propositions. The next proposition
assures the gadgets produced by $R$ and used in our reduction are available in
uniformly random graphs.

\BP\label {p.dom} The probabilities of graphs $G$ output by $R$
 (from $\l$-random $X,\a$) are $O(\|\a\|\l(\{G\}))$. \EP

\BPR Assigning edges between successive nodes in $U_T$ to encode {\em random}
RRTP instance agrees with $\l$. Thus we only need to show that other
constraints $R$ enforces are satisfied by a sufficient fraction of graphs.

By Corollary~\ref {c.max} and Proposition~\ref {p.codes}, the set $H$ of graphs
with a $k$-node tournament $T$ have probability $\l_c(H)\sim2^{kc}\l(H)=\l(H)
\cdot O(n)$. With such $T$, the probability of sizes of $L_T$ and $U_T$ being
their expected value $2p^2$ is $1/O(n)$ by item 2(c) of Remark~\ref {l-prob}.
All other conditional probabilities are $1-o(1)$, except for four $t_1$ links
being double, which is $1/81$.\EPR

\BP\label{p.3} With all $X$ and a.e.\@ $\a$, the reduction $R$ succeeds on
$Y=(X,\a)$ i.e., $G=R(Y)$ has a $C$-coloring $w$ if and only if $X$ has a
tiling $w'$. Moreover, $w$ is constructed in P-time (by $Q(w)$). \EP

The proof is the topic of the next section. It describes the standard
coloration $C$ required by $R$ and the inverse $Q^{-1}$ of $Q$ mapping tiling
solutions to $C$-colored graphs. The crucial part is to assure item~\ref
{red-Q} of the definition of reduction: that $Q$ can transform any witness of
coloring problem into a witness of tiling.

 \section {The Coloration C and the Proof of Proposition~\ref{p.3}}
 \label {sec-grid}

\pp {Enforcing Structures with Spots; Bootstrapping.} We now give simple and
typical examples of how our design of $C$, restricting spots allowed in the
graph (called $C$-spots), is used. Properties that all $C$-colored graphs have
we call \trm {forced by} $C$. For instance, our $C$ has only one spot with 3
blank (blank-looped) nodes and it is a tournament. This forces all blank nodes
of any $C$-colored graph to form a tournament as well.

Another example: no $C$-spots have nodes with two links of certain types, say,
outgoing red edges without reverse. This forces the same on any $C$-colored
graph. Yet, local conditions alone, such as those imposed by spots, cannot
ensure all needed features, e.g., at least one link of a given type at certain
nodes. But our $C$ also requires leaving enough edges blank. We show this
assures the maximal size of the above tournament $T$ (and thus its unique
location, bootstrapping our construction). Indeed, some links with blank edges
are only at blank nodes, and any smaller $T$ lowers the number of blank edges
obtainable by using only $C$-spots.

Next we describe a colored graph template reflecting any solved Tiling instance
and random graphs matching its input row. We force it to be displayed by any
adversary $C$-coloring such a random graph.

\pp {Mapping Tiling to Colored Template.} We first make blank the loops of $T$
and make blank-yellow all double (i.e., double-edge) links between $L$ and $T$.
We then implement $Q^{-1}$ as outlined in Remark~\ref {rem-three}. First, we
transform the tiling into a colored ``template'' graph $G'$. Then we di-embed
$G'\setminus G$ into $G$, copying the colors. All other edges in $G$ will be
yellow. $C$ will assure any coloring of G to have this form. See figure
\ref{trans}.

Our $G'$ consists of $T,U_T$ and, in addition, a toroidal \trm {grid} -- a
product graph $\Z/2p\times\Z/p$, including $z_0=t_1$ located at the \trm
{origin} of the grid. Grids's smallest \trm {squares} (undirected 4-node
cycles) represent RRTP tiles and are ordered by rows and by columns in each
row. The corners of each $i$'th square are connected by \trm {radial} links
into its \trm {center} $v_i{\in}U_T$ (treated thereafter as a part of the
grid). We also need the random edges of the graph $G$ to determine uniquely the
coloring of the grid's \trm {floor row} containing $z_1$. For that we include
in $G'$ its code links to $T$. They assure (sec.~\ref {sec-mono}) the correct
placement of $v_i$ centers in the first row:

\begin{enumerate}
 \itemsep2pt\parsep1pt
 \item The codes of section~\ref {R} determine a monotone
       order of $v_i$ nodes forming the \trm {input chain}.
 \item The coloring of the code edges proves the correctness of this ordering.
 \item The connections of the input chain consecutive nodes
       determine the coloring of the grid's floor row.
 \item The coloring of the rest of the grid exhibits
       a solution to the tiling problem instance.\end{enumerate}

We may denote colors as R (red) G (green) Y (yellow) and loosely refer to the
absence or blankness of an edge as to two special ``colors'' (A and B). An edge
colored red (resp., green, etc., or left blank), we call a red edge (resp.,
green, blank, etc.); same for double edges, e.g., a yellow edge with the
reverse blank edge is yellow-blank. The node's color (or blankness) is that of
its loop. We denote the induced colored 3-node and 2-node subgraphs (spots and
links) as $[u,v,w]$ and $[v,t]$, respectively. Our grid's L-nodes have two
vertical $\vl,\vl^{-1}$, two horizontal {\hl} and four radial
{\il,\jl,\kl,\lln} links each. The {\hl} links are directed \trm {idle} or
undirected: \trm {active} and \trm {wall}. We artificially refer to active
links as \trm {outgoing} and wall as \trm {incoming} at both ends.

With these convention each grid node has exactly one link of each type
\il,\jl,\kl,\lln, and L-nodes one incoming and one outgoing {\hl} and {\vl}.
So, we can treat them as functions: if $\vl$ connect $x$ to $y$ and $\il$
connects $y$ to $z$, we write $y=\vl(x), z=\il(y)=\il\vl(x)$.
 Except for non-idle \hl, we also can use inverses, $x=\vl^{-1}(y)$.
 We define \hl' as $j.i$ and will prove this permutations to commute with \vl.
 The column of the origin $z_1$, we call the \trm {wall}.

By \ic-links we call {\il} in $L_T$, {\kl} from $U_T$ to the wall, or {\jl} at
the rest of $U_T$. They link our input chain $z_1,v_1,z_2,v_2,z_3,\ldots,
z_{2p^2}, v_{2p^2},\, z_i\ne z_j\text { if }i\ne j$. It passes through all $2p$
floor nodes, descends to the wall of the next lower rows via $\kl$, and
eventually passes through all the grid nodes.

The coloring of {\vl} links represents a solution to the Tiling Problem.
 This includes a direction bit \lr
 \\
 (also in \il,\lln) used below. {\jl} at $z_1$ carries wall {\wl} and floor
{\fl} flags; they propagate: {\wl} via {\jl,\lln} chain, {\fl} via {\jl,\il}
chain. More details are in section~\ref {sec-ts}. Then we use the key fact that
$G'\setminus T$ {\em has degree} $O(1)$. This allows to prove in section~\ref
{sec-ts} the existence of an identical on $T\cup U_T$ di-embedding $g_{\sX}$ of
$G'$ into a.e.\@ $G$.

\subsection {Forcing the Blank Tournament} \label {sec-blanks}

This section shows how a selection of spots with blank edges enforces a
$\Z/2p\times\Z/p$ grid, the tournament and the input chain. First, we show that
coloring a tournament of size $<k$, forces, in a.e.\@ $G$, a $\Theta(n)$
shortage of blank edges (between $L$ and $T$); this cannot be compensated by
coloring edges of other types which are $o(n)$ altogether. Since $T$ is the
only $k$-node tournament, the $C$-coloring must use it. Now, each $C'$-link
with blank edges can occur at most $b_i$ times in a.e.\@ random graph.
Requiring total $\sum b_i$ blank edges forces the coloring to use each type
exactly $b_i$ times. This strategy crucially depends on $G$ being random.

Let the graph $G$ produced by the reduction $R$ be $C$-colored. Let $\widetilde
T$, $L_{\widetilde T}$, and $U_{\widetilde T}$ be respectively its sets of
blank nodes, of looped (non-blank) and of unlooped nodes connected to every
node in $\widetilde T$.

Any link with a blank non-loop edge falls into one of the following disjoint
{\em directed} types: $\vl,\jl,\il,\key,\dwn$.

 The following constraints \ref{A}-\ref{D}
 on these types are enforced by the spots in $C'$.

\begin {enumerate}
 \itemsep4pt\parsep2pt
 \renewcommand\labelenumi{(\roman{enumi})}\renewcommand\theenumi\labelenumi
 \item\label{A} Each 3-node induced subgraph of $\widetilde T$
 (and thus the whole $\widetilde T$) is a tournament;
 let $\widetilde t_1$ be its sink node.
 \item\label{B} Non-blank nodes have at most one link of each of
     the types: \key,\il,\jl,\kl,\lln, incoming and outgoing \vl,\hl.
 \item\label{C} All {\il,\jl,\kl,\lln} run from $U_{\widetilde T}$
      to $L_{\widetilde T}\cup \{\widetilde t_1\}$, {\vl,\hl}
      within $L_{\widetilde T}\cup{\{\widetilde t_1}\}$, {\key}
      single B or Y-B from $U_{\widetilde T}$ to $\widetilde T$.
 \item\label{D} {\dwn}-links are $L-T$ doubles with a blank edge to $T$
 (in two places it is from $T$).
 \end{enumerate}

Put $b_0{=}k$, $b_1=b_2=b_3=b_4=2p^2$, $b_5{=}$ the number of double edges $L$
to $T$, and $b{=}\sum b_i$. Let $\widetilde b_0$ be the number of blanks on
$\widetilde T$ and $\widetilde b_1,\widetilde b_2,\ldots, \widetilde b_5$ be
the number of edges of types $\vl,\il,\jl,\key$, and $\dwn$, respectively. Let
$\widetilde b=\sum_i\widetilde b_i$ be the number of all blank edges.

\BP\label{rem-max} Any coloring, respecting \ref{A}-\ref{D} and $\widetilde
b\ge b$, of a.e.\@ $G\in\SZ$ has $\widetilde b_i=b_i$ and $\widetilde T=T$.\EP

\BPR By \ref{A}-\ref{D}, no non-$T$ node has ${>}2$ outgoing blank non-{\dwn}
edges.\\ Let $|\widetilde T|{=}k'{=}k{-}\del$. Then, in a.e.\@ $G$,

\begin{enumerate} \item $\del\ge 0$, since by Corollary~\ref {c.max} $k'\le k$.
 \item $<(3/4)^{k'}2n$ nodes are connected to all nodes of any given
 $k'$-node set $B$ (by Remark~\ref{l-prob}-\ref{r22b}). \end{enumerate}

Any coloring of $G\in\SZ$ has $b_5-\widetilde b_5>\del n/9$. To have
$\widetilde b\ge b$, one has to compensate by extra blank edges of other types,
whose total number is $<4(\frac34)^{k'}n$. This is possible only if
$4(\frac34)^{k'}n> \frac{\del n}9$ i.e., $(\frac43)^{\del}>
(\frac43)^k\cdot\frac {\del}{36}$, whence $\del{>}k$ for large enough $k$,
which is a contradiction. It follows that $|\widetilde T|{=}k$, and by
uniqueness of the $k$-node tournament, $\widetilde T{=}T$. Thus $U_{\widetilde
T}{=}U_T$, and $L_{\widetilde T}{=}L_T$. The number of blank edges of each type
$\vl,\il,\jl,\key$, by \ref {C} is at most $b_1=b_2=b_3=b_4=2p^2$. Finally, the
number of loops on $T$ is $k$. Thus, $\widetilde b_i\le b_i$ as claimed.\EPR

\subsection {Forcing of a Monotone Order on the Input Chain and of the Toroidal
Grid}\label {sec-mono}

We exhibit the decreasing codes order of $v_i$'s as follows. For each $i$, we
choose $l(i)$ and color a \key-link $(v_i,t_{l(i)})$. The existence of $b_4$
$\key$-edges implies that every $v_i{\in} U_T$ has a \key-link. The conditions
described next imply that the colors of links from $v_i,z_{i+1},v_{i+1}$ to $T$
indicate that the code of $v_{i+1}$ w.r.t. $t_j$ is at most that of $v_i$ for
$j>l(i)$ (highest bits), and strictly less for $j=l(i)$. Thus, the input nodes
form an acyclic \ic-linked path in the decreasing order of their codes. Note
that $x>y$ as integers represented as ternary $k$-digit strings, iff for some
$l\le k$ and all $j>l$ digits $x_j\ge y_j$ and $x_l>y_l$. Recall that a node in
$U_T\cup L_T$ is connected to every node in $T$. The code of $v_i$ w.r.t. $t_j$
is copied onto $[z_{i+1},t_j]$ and will be compared in $[z_{i+1},t_j,v_{i+1}]$.

\pp {Coloring the code links.} All edges $t_j$ to $v_i$ and \trm {down} edges
(single edge to $t_j$) from $v_i$ are: red if not down and $j{>}l(i)$, blank if
$j{=}l(i)$, or yellow otherwise. The code of $v_i$ w.r.t. $t_j$ is
$\mythree,\myone$, or $\mytwo$ to reflect $v_i$ having a down, up, or double
edge to $t_j$. For $j>l(i)$, this ternary code is exhibited on the
$z_{i+1}{-}t_j$ link by coloring its edges yellow, red, or green respectively.
For $j=l(i)$, $\mytwo$ is encoded on the $z_{i+1}{-}t_j$ link if $v_i$ has a
code $\mythree$ w.r.t. $t_j$, otherwise $\myone$. For $j{<}l(i)$, the yellow
$z_{i+1}{-}t_j$ link encodes an $\mythree$. Now we can, in
$[z_{i+1},v_{i+1},t_j]$, uniformly for all $i,j$, prohibit the codes of
$v_{i+1}$ to exceed those encoded at $z_{i+1}$.

\BPR [Toroidal Grid.] First, we prove the group $\Gamma$ generated by
permutations \vl,\hl' is commutative. Indeed, consider a \vl\il-path $x{-}o$
(a \vl-link from $x$ followed by an \il-link to $o$), a $\jl\vl^{-1}$-path
$o{-}y$, and an $\il\jl$-path $x{-}z$. By our $C'$, they will require an
\lln-link $x{-}o$, a \kl-link $o{-}y$, and an \hl-link $x{-}z$; the
\lln\kl-path requires an \hl-link $x{-}y$ and {\il,\lln} carry the same
direction bit \lr. Depending on \lr, the \hl-links at $x$ are both \trm
{outgoing} or both \trm {incoming}. By $C'$, $x$ cannot have two such links,
thus $y=z$, $\hl'=\vl\il\jl\vl^{-1}$, so $\hl'\vl=\vl\il\jl=\vl\hl'$.\EPR

\pp {Claim:} \vl,\hl' links induce a connected toroidal grid on $V$.

\BPR \ic-chain spans the whole grid $V{:}{=}L_T\cup U_T\cup\{z_1\}$.
Thus $z_1\Gamma=L_T$: any $u\in L_T$ is accessible from $z_1$ as
$u=\hl'^i\vl^j(z_1)$. Now, $\hl'^r(z_1)=\vl^s(z_1)=z_1$ for some minimal $r,s$,
since \vl,\hl' permute the finite $V$. If $\hl'^i\vl^j(z_1)=z_1$ then
$\hl'^i(z_1)=\vl^j(z_1) =z_1$ and $r|i$, $s|j$, since blank $z_1$ is the only
node with both {\wl,\fl} flags up in \jl. Then, the $\hl'^i\vl^j(z_1)
\leftrightarrow(i,j)$ bijection sets up a rectangular grid co-ordinates on $V$
and $rs=2p^2$. By our $C'$, $r,s>2$, so the grid is either $2p\times p$ or
$p\times 2p$. In our (toroidal) RRTP, exactly one $\il$-link changes direction
at each row of any tiling. Hence $s$ is even and $s{=}2r{=}2p$.\EPR

\subsection {Forcing the Correct Representation of Tiling} \label {sec-ts}

 \begin{figure}
 [htb]
 \setlength{\unitlength}
 {1pt}
 \centering\noindent\begin{picture} (450,300)(-10,-6)
 \newcommand\putbx[2]{{\put(#1){\makebox(0,0){\fbox{\emm{#2}}}}}}
 \newcommand\putm [2]{{\put(#1){\makebox(0,0){\emm{#2}}}}}
 \put (0,0) {\line (1,0) {432}} \put (0,0) {\line (0,1) {288}}
 \put (0,0) {\line (1,1) {288}} \put (432,288) {\line (-1,-1) {288}}
 \put (432,288) {\line (-1,0) {432}} \put (432,288) {\line (0,-1) {288}}
 \put (0,288) {\line (1,-1) {288}} \put (432,0) {\line (-1,1) {288}}
 \put (0,144) {\line (1,0) {432}} \put (288,0) {\line (0,1) {288}}
 \put (144,0) {\line (0,1) {288}} \put (0,144) {\line (1,1) {144}}
 \put (0,144) {\line (1,-1) {144}} \put (432,144) {\line (-1,1) {144}}
 \put (432,144) {\line (-1,-1) {144}}
 \put (0,0) {\circle {25}} \put (144,0) {\circle {25}}
 \put (288,0) {\circle {25}} \put (432,0) {\circle {25}}
 \put (0,144) {\circle {25}} \put (144,144) {\circle {25}}
 \put (288,144) {\circle {25}} \put (432,144) {\circle {25}}
 \put (0,288) {\circle {25}} \put (144,288) {\circle {25}}
 \put (288,288) {\circle {25}} \put (432,288) {\circle {25}}
 \put (72,72) {\circle* {5}} \put (216,72) {\circle* {5}}
 \put (360,72) {\circle* {5}} \put (72,216) {\circle* {5}}
 \put (216,216) {\circle* {5}} \put (360,216) {\circle* {5}}
 \putm {144,294} {\trit_C} \putm {288,294} {\trit_D}
 \putm {159,282} {C} \putm {303,282} {D}
 \putm {36,249} {(\rht,\fl)} \putm {108,252} {(\fl,\wl)}
 \putm {180,252} {(\lr_C\,,\,\fl\,)} \putm {252,252} {(\fl,\wl)}
 \putm {324,252} {(\lr_D\,,\,\fl\,)} \putm {396,252} {(\fl,\wl)}
 \putm {4,216} {(\rht,\bit_L)} \putm {146,216} {(\rht,\bit_C)}
 \putm {291,216} {(\lft,\bit_D)} \putm {428,216} {(\lft,\bit_R)}
 \putm {72,207} {E} \putm {216,207} {F} \putm {360,207} {G}
 \putm {36,180} {(\rht,\rht)} \putm {108,180} {(\wl)}
 \putm {180,180} {(\rht,\lr_C)} \putm {252,180} {(\wl)}
 \putm {324,180} {(\lft,\lr_D)} \putm {396,180} {(\bit_R)}
 \putm {72,150} {\rht} \putm {144,150} {\trit_A}
 \putm {216,150} {(\lr^-,\bit_A,\bit_B)}
 \putm {288,150} {\trit_B} \putm {360,150} {\lft}
 \putm {-6,144} {L} \putm {158,138} {A} \putm {303,138} {B} \putm {438,144} {R}
 \putm {33,108} {(\rht,\fl)} \putm {110,108} {(\fl,\wl)}
 \putm {177,108} {(\rht,\fl)} \putm {252,108} {(\fl,\wl)}
 \putm {324,108} {(\lft,\fl)} \putm {396,108} {(\fl,\wl)}
 \putm {144,72} {(\lr^-,\bit_A)} \putm {288,72} {(\lr^-,\bit_B)}
 \putm {428,72} {(\lft,\bit_R)}
 \putm {72,63} {O} \putm {216,63} {P} \putm {360,63} {Q}
 \putbx {47,85} {\il} \putbx {96,85} {\jl} \putbx {8,72} {\vl}
 \putbx {96,35} {\kl} \putbx {47,35} {\lln} \putbx {72,8} {\hl}
 \end{picture}\caption {Simulation of a transition.}\label{trans}\end{figure}
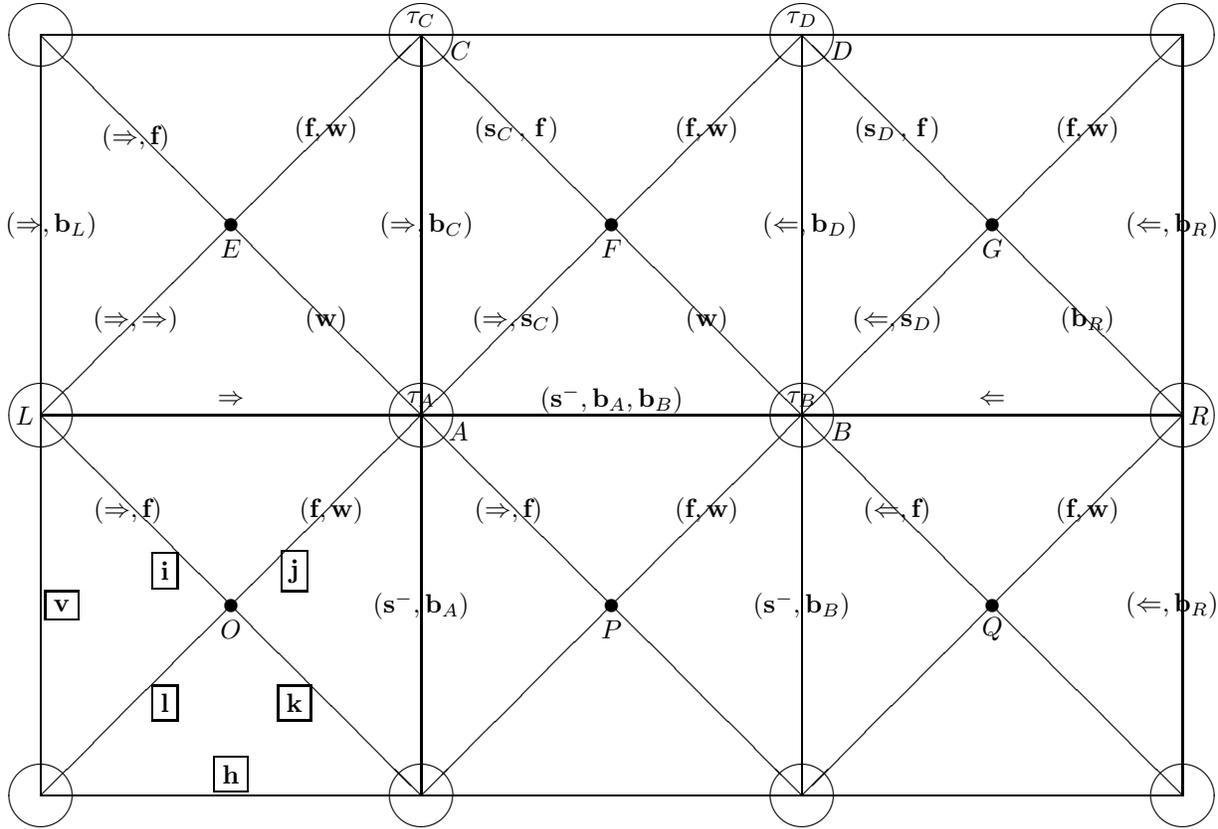

\pp {Link Bits.} The grid links (see figure~\ref {trans}) reflect RRTP fields:
trit $\trit$, bit $\bit$, side $\lr$, floor $\fl$, and wall $\wl$. The
self-loop color at a node in $L_T$ carries its $\trit$; now we list the bits
carried by its incident links ($+/-$ refers to their
one-transition-later/earlier values): $\vl\!:\bit+,\lr$; (i.e.$\vl$-links carry
$\bit+,\lr$) $\il\!: \lr,\fl$; $\jl\!: \fl,\wl$; $\lln\!: \lr,\lr+$; $\kl\!:
\bit$ if idle, else $\wl$; $\hl\!: \lr^-,\bit$ of both active ends, or $\lr^-$
of an idle end. The same triple $(\trit,\bit,\lr)$ can encode a state (if
$\lr{\ne}\lr^-$) or a symbol (if $\lr{=}\lr^-$).

\pp {Link Coloring.} Blanks and loops distinguish all link types (except {\kl}
from {\lln} that together have 12 edge patterns); {\vl,\il,\jl} each have a
blank edge with (R/G/Y/A) reverse. $\hl$-links have no blank edges. Each row of
the torus has one active and one wall \hl-link connected by two chains of idle
links. Idle {\hl} have a red edge directed toward the active \hl, and a green
reverse. Active links carry $(\lr^-,\bit_1 ,\bit_2)$ and have a red (iff
$\lr^-{=} \lft$) or else a green. If $\bit_1\ne\bit_2$, {\hl} has a yellow edge
pointing to $\bit=1$. If $\bit_1=\bit_2$, {\hl} is monochromatic, double iff
$\bit=1$, else rightward single. If red single with red loops, it is the wall
\hl-link.

\pp {Link Chains.} The {\il-\jl} and {\jl-\kl} chains propagate the floor and
the wall flags respectively from the origin $z_1$ to its entire row and column.
The wall nodes (in $z_1$'s column) represent the symbol D of RRTP, their left
neighbors -- the symbol ``0''. They are oriented ``back-to-back'' and connected
by a special wall \hl-link. It requires the wall-flag up at {\jl,\kl} links and
vice versa. Adjacent idle \hl-links match directions. Thus, chains of leftward
pointing and of rightward pointing idle \hl-links can only meet at a {\em
unique} active \hl-link.

\pp {RRTP Compliance: the Base Case.} We now tell how $C$-spots force copying
the RRTP input onto the grid's floor row (the reduction algorithm $R$,
Sec.\ref{R}, step \ref{P4}). On the floor's \jl-\il-triangles (e.g., PBQ in
fig.~\ref {trans}), the presence of edges between unlooped ends (PQ) is copied
onto its G/R/Y trit (e.g., $\trit_B$). Floor's \il-link at active B has a
Y-loop (used for the $*$-symbol and for the starting $e$-state). This precludes
edges between $P$ and $Q$. This is the leftmost pair of disconnected successive
input nodes on the \ic-chain: others to the left are blocked by rightward
\hl-links (via rightward \il-links). The left (rightward directed) segment has
trits symbols from $\{0',1',*'\}$, it ends at the active \hl-link, whose right
node encodes the starting state. The right segment acts similarly except that
only the rightward edge $(P,Q)$ is used and mapped into R/G trit colors.

\pp {Induction on Rows.} In the figure, nodes $A$ and $B$ are active, their
incident edges carry the RRTP triples $(\trit_A,\lr_A,\bit_A)$, $(\trit_B,
\lr_B,\bit_B)$. Since $A$ and $B$ have opposite $\lr$ values in the current
step, the incoming $\lr^-$ are the same so that $\lr_A^-=\lr_B^-=:\lr^-$ as
shown in the figure. Now, the $\hl$-link carries $(\lr^-,\bit_A,\bit_B)$. The
triangle $ABC$ with the $\hl$-link on $AB$ determines $\bit_C$ on the
$\vl$-link $AC$ and $\trit_C$ at $C$. The triangle $ABF$ imposes $\lr_C$ on the
$\lln$-link $AF$. Node $D$ transition is similar. This assures the
representations of that tile's top symbols at the next row. The $\il$-links
$FC$ and $GD$ copy $\lr_C$ and $\lr_D$ respectively. $R$ is idle so its
$\vl$-link's $\bit_R$ is copied from the incoming $\vl$-link (via \kl-link).
This structure of computation and data flow is similar in each square.

\BPR [Proof of Proposition~\ref {p.3}.] Recall that a.e.\@ graph $G$ sampled by
the reduction $R(X,\a)$ is in $\SZ$, {i.e.} has a unique $k$-node tournament
$T$, $|U_T|=|L_T|{+}1=2p^2$, and distinct $U_T$ codes. The tiling instance $X$
determines the edges between successive input nodes. We have described the
graph $G'$ and its $C$-coloration that must encode the tiling pattern. The
degree of $G'\setminus T$ is $O(1)$. Thus, by the Embedding Lemma~\ref {l.emb},
$G'$ can be di-embedded in a.e.\@ $G$, {i.e.} ${\bar\l}(\exists
g:G'\hookrightarrow G) \sim1$.\EPR

\pp {Acknowledgments.}
 \hspace*{-10pt}
 We thank Peter G\'{a}cs for discussions and
 Marvin Minsky for comments on small UTMs.

 \end {document}